\documentclass[letterpaper,12pt]{article}
\usepackage{color,amssymb,enumerate,float}

\usepackage{ifpdf}
\ifpdf
	\usepackage[pdftex]{graphicx}
	\usepackage[pdftex,unicode,implicit]{hyperref}

	\hypersetup{
  	pdftitle     = {}, 
  	pdfkeywords  = {},
  	pdfauthor    = {},
  	pdfcreator   = {pdf\LaTeXe\ with package \flqq hyperref\frqq},
  	pdfproducer  = {pdf\LaTeXe\ with package \flqq hyperref\frqq},
  	pdfpagemode  = UseNone,  
  	pdffitwindow = true,  
  	unicode      = true,
  	plainpages   = true,
  	colorlinks   = true,  
  	citecolor    = black,  
  	urlcolor     = blue, 
  	linkcolor    = black
	}

\else

  \usepackage[dvips]{graphicx}

\fi 

\makeatletter
\@addtoreset{equation}{section}
\makeatother

\begin{document}
	
\thispagestyle{empty}

\begin{center}
{\bf \LARGE The point-particle solution and the asymptotic flatness in $2+1$ Ho\v{r}ava gravity}
\vspace*{15mm}

{\large Jorge Bellor\'{\i}n}$^{1}$
{\large and Byron Droguett}$^{2}$
\vspace{3ex}

{\it Department of Physics, Universidad de Antofagasta, 1240000 Antofagasta, Chile.}
\vspace{3ex}

$^1${\tt jbellori@gmail.com,} \hspace{1em}
$^2${\tt byron.droguett@ua.cl}

\vspace*{15mm}
{\bf Abstract}
\begin{quotation}{\small\noindent
  We show that the solution corresponding to the gravitational field of a point particle at rest in $2+1$ nonprojectable Ho\v{r}ava is exactly the same as $2+1$ General Relativity with the same source. In General Relativity this solution is well known, it is a flat cone whose deficit angle is proportional to the mass of the particle. To establish the system we couple the Ho\v{r}ava theory to a point particle with relativistic action. Motivated by this solution, we postulate the condition of asymptotic flatness exactly in the same way of $2+1$ General Relativity. A remarkable feature of this condition is that the dominant mode is not fixed, but affected by the mass of the configuration. In this scheme, another important coincidence with $2+1$ General Relativity under asymptotic flatness is that the energy is the same (except for some coupling constants involved), the $z=1$ term with the derivative of the lapse function does not contribute.
}
\end{quotation}

\end{center}

\thispagestyle{empty}

\newpage
\section{Introduction}
In gravitational physics asymptotic flatness is a key concept in the study of finite-size, localized systems. In simple grounds, one wants a way of getting arbitrarily far away from the localized system. In $3+1$ and $2+1$ dimensional General Relativity asymptotic flatness has well-established definitions. Interestingly, the scenario in the $2+1$ theory is substantially different to the $3+1$ case since the $2+1$ definition is not based on having a fixed metric at infinity \cite{Ashtekar:1993ds}. The dominant mode in the asymptotic expansion varies functionally among the admissible configurations. This can be traced back to the variability of the mass originating the asymptotically flat configuration. In general, $2+1$ General Relativity is quite different to its four-dimensional counterpart since it is a topological theory, in the sense that the vacuum theory without cosmological constant has vanishing curvature, whereas with cosmological constant it yields constant curvature. Despite this, the theory is far from being trivial, as it is exemplified by the well known case of the Ba\~nados-Teitelboim-Zanelli black hole solution with a negative cosmological constant \cite{Banados:1992wn}. Further developments of the asymptotic flatness and the energy in $2+1$ General Relativity can be found, for example, in Refs.~\cite{Marolf:2006xj,Corichi:2015nea,Miskovic:2016mvs}.

On the other hand, our understanding of the gravitational interactions is incomplete since there is not a definitive gravitational theory valid at quantum scales. Ho\v{r}ava theory \cite{Horava:2009uw,Horava:2008ih} is precisely a proposal for the consistent quantization of the gravitational interactions. It is based on a foliation of spacelike hypersurfaces endowed with absolute physical meaning. In consequence, the allowed symmetry corresponds to those coordinate transformations that preserve the preferred foliation. The theory is made with higher order spatial-derivative terms, defining in this way a power-counting renormalizable theory. Indeed, the complete renormalization of the projectable version, in which case the lapse function is restricted to be a function only of time, has been proven \cite{Barvinsky:2015kil}.

The generic formulation of the Ho\v{r}ava gravity propagates one physical degree of freedom in addition to the two tensorial modes that are also present in four-dimensional General Relativity. This holds excluding the cases of the kinetic-conformal formulation \cite{Bellorin:2013zbp}, which is defined at a critical point of the theory, and other modifications like models with further gauge symmetries \cite{Horava:2010zj}. In these cases the extra mode is eliminated. In the generic formulation the extra mode exhibits the so-called strong-coupling problem, which can be circumvented by lowering the scale of activation of the higher order operators \cite{Charmousis:2009tc,Papazoglou:2009fj,Blas:2009ck}.

The scalar mode is also present in the $2+1$ dimensional Ho\v{r}ava theory, where it is the unique propagating mode. A clear identification of this mode in the nonprojectable $2+1$ Ho\v{r}ava theory has been carried out in Ref.~\cite{Sotiriou:2011dr}. The presence of this mode brings an interesting scenario since the $2+1$ dimensional theory becomes nontopological. Thus, Ho\v{r}ava theory provides us a truly local gravitational theory in $2+1$ dimensions. It contains a physical mode candidate for carrying the gravitational interaction in the quantum formulation, even in the vacuum and in the absence of cosmological constant.

A better understanding of the $2+1$ Ho\v{r}ava theory requires a proper definition of the asymptotic-flatness condition necessary for localized systems. In this paper we address this problem for the generic nonprojectable version of the Ho\v{r}ava theory, excluding the case of the kinetic-conformal critical point. To define the asymptotic-flatness condition, we adopt the same criterium used in $2+1$ General Relativity \cite{Ashtekar:1993ds}. In $2+1$ General Relativity the exact solution corresponding to the gravitational field of a point particle at rest is known \cite{Deser:1983tn}. Excluding some ranges of values for the mass of the particle, the solution corresponds to a conical singularity with a flat space outside the source. The conical angle depends on the mass. Based on this, the criterium for defining an asymptotically flat configuration is established as the condition that such a configuration must approach the point-particle solution for large enough distances. The fact that the conical angle of the solution depends on the mass translates to the definition of the asymptotic flatness, having as a consequence that the leading mode of the expansion is not fixed, but functionally variable, unlike the asymptotically-flat condition in four dimensional gravity.

In this paper we carry out the same two main steps in the $2+1$ Ho\v{r}ava theory: first, we search for the point-particle solution and, second, we postulate the definition of the asymptotic flatness based on this solution. We start by setting the coupling of the Ho\v{r}ava theory to the particle. We keep a relativistic dynamics for the particle, but living in a ambient space governed by the Ho\v{r}ava theory. Therefore, we evaluate the relativistic line element of the particle on a geometrical background given by the Arnowitt-Deser-Misner (ADM) field variables of the Ho\v{r}ava theory. The motivation for this choice is the fact that relativistic physics is strongly supported by tests in the matter sector, hence we wish to keep our model, although three-dimensional, as realistic as possible. Of course, other choices are possible. With regard this we comment that we obtain the gravitational solution with the particle at rest. The effect of a particle at rest in the gravitational field equations can be embedded in several particle dynamics, relativistic or not, contributing in the same way by means of a delta-source term. In addition, we consider the coupling of the particle to the large-distance effective action of the Ho\v{r}ava gravity, which is given by the lowest order terms in derivatives in the potential. This is a good scenario for studying asymptotic flatness, where gravitational fields are supposed to be weak. We remark that the effective action is foliation-preserving-diffeomorphisms (FDiff) invariant, hence nonrelativistic. In particular it includes the $(\partial_i \ln N)^2$ term in the potential, where $N$ is the lapse function and $\partial_i$ are spatial derivatives.

We give in advance the rather surprising result that the point-particle field solution is exactly the same as in General Relativity (except for different coupling constants). As part of our ansatz we consider static lapse function and static spatial metric, since a static gravitational field is the natural setting for a particle at rest. An immediate consequence of this is that the condition of asymptotic flatness that we postulate in 2+1 nonprojectable Ho\v{r}ava theory is the same as in $2+1$ General Relativity. 

As it is well known, the definition of asymptotic flatness has consequences on the differentiability of functionals like the Hamiltonian \cite{Regge:1974zd}, which in turns connects with the value of the energy for this kind of configurations. The need of a differentiable Hamiltonian for asymptotically flat configurations adds a boundary counterterm, which is the ADM energy in the $3+1$ case. A similar relationship occurs in $3+1$ nonprojectable Ho\v{r}ava theory \cite{Donnelly:2011df,Bellorin:2011ff}, where the gradient flux of the lapse function at infinity contributes with the energy, besides the ADM term. This happens also in the kinetic-conformal formulation, whose canonical version is obtained independently of the generic case due to the extra constraints \cite{Bellorin:2013zbp}. Simultaneously to the definition of the asymptotic-flatness condition in $2+1$ General Relativity, in Ref.~\cite{Ashtekar:1993ds} the boundary counterterm giving the energy was found by means of variations in the space of $2+1$ asymptotically flat configurations. Since in this paper we undertake the issue of the $2+1$ asymptotic-flatness in Ho\v{r}ava theory, an natural subsequent step is to apply it to discuss the differentiability of the Hamiltonian, searching for possible counterterms giving the energy for asymptotically flat configurations. We anticipate again the result: it turns out that the energy is the same of $2+1$ General Relativity. This is also surprising and quite different to the $3+1$ Ho\v{r}ava theory since we find that there is no contribution of the flux of the lapse function at the infinite boundary.

This paper is organized as follows. In section 2 we present the large-distance effective nonprojectable $2+1$ Ho\v{r}ava theory coupled to the relativistic particle. In section 3 we find the solution for the gravitational field of the particle at rest. In section 4 we define the condition of asymptotic flatness, which is the same of the $2+1$ dimensional General Relativity. In section 5 we perform the canonical formulation of the $2+1$ theory, with the final aim of studying the differentiability of the Hamiltonian and the boundary terms giving the energy.

\section{Lagrangian Formulation}
As we have commented, our strategy for posing the system Ho\v{r}ava gravity-point particle is to consider a relativistic particle moving on a (dynamical) background governed by the field equations of the Ho\v{r}ava theory. The action of the relativistic particle is the usual embedding of the 4-dimensional length of its trajectory into the ambient spacetime. In Ho\v{r}ava theory the analogous role of spacetime is played by the ADM variables, hence we may use them to build the background on which the particle moves. The spacetime metric can be build in terms of the ADM variables according to
\begin{equation}	
g_{\mu\nu} =
\left( {\begin{array}{cc}
	-N^{2}+N_{k}N^{k} & N_j\\
	N_i & g_{ij} \\
	\end{array} } \right) \,,
\end{equation}
where $N$ is the lapse function, $N_{k}$ is the shift function and $g_{ij}$ is the spatial metric (spatial indices are raised and lowered with $g_{ij}$). To simplify the discussion we choose the time coordinate $t$ of the ambient foliation to parametrize the world line of the particle. The mechanics of the particle is characterized by the embedding fields $q^0 = q^0(t)$ and $q^i = q^i(t)$, which define the position of the particle in the foliation. Thus, whenever they are involved in the particle dynamics, the ADM variables are evaluated on the position of the particle, $g_{ij}(q(t))$, and so on. On the side of the Ho\v{r}ava gravity we consider the effective theory for large distances, which is given by the potential of second order in spatial derivatives,
\begin{equation}
 \mathcal{V} = - \beta R - \alpha a_k a^k \,,
\end{equation}
where $\beta$ and $\alpha$ are coupling constants and
\begin{equation}
 a_i = \frac{ \partial_i N }{ N } 
\end{equation}
is a FDiff-covariant vector \cite{Blas:2009qj}. Since we are interested in asymptotic flatness we do not consider a cosmological-constant term in the potential. The combined system Ho\v{r}ava gravity-point particle in $d$ spatial dimensions is given by the action
\begin{equation}
	S = 
	\frac{1}{2\kappa} \int dt d^dx 
	\sqrt{g} N \left( K_{ij}K^{ij} - \lambda K^2 
	+ \beta R + \alpha a_k a^k \right) 
	- m \int dt \sqrt{ L } \,,
	\label{action}
\end{equation}
where  
\begin{eqnarray}
 &&
 K_{ij} = 
 \frac{1}{2N} \left( \dot{g}_{ij} - 2 \nabla_{(i} N_{j)} \right) \,,
 \label{extrinsiccurvature}
 \\ && 
 L = 
 (N^{2} - N_{k}N^{k}) \left( \dot{q}^0 \right)^2 
 - 2 N_k \dot{q}^0 \dot{q}^k - g_{kl} \dot{q}^k \dot{q}^l \,,
 \label{L}
\end{eqnarray} 
$K \equiv g^{ij} K_{ij}$, $m$ is the mass of the particle, $\kappa$ and $\lambda$ are coupling constants and the dot stands for derivative with respect to $t$, $\dot{g}_{ij} \equiv \partial g_{ij} / \partial t$. The tensor (\ref{extrinsiccurvature}) is the extrinsic curvature tensor of the leaves of the foliation. $L$ is the squared line element of the particle evaluated on the background of the ADM variables, and these variables are evaluated at the position of the particle in $L$. 

The field equations of the ADM variables are obtained by varying the action (\ref{action}) with respect to them. This yields
\begin{eqnarray}
	&& 
	 K^{ij}K_{ij} - \lambda K^{2} + \beta R + \alpha a_i a^i  
     - 2 \alpha \frac{\nabla^{2} N}{N} 
	 =
	 2 \kappa m \frac{ (\dot{q}^0)^2 N }{ \sqrt{gL} } 
	   \delta^{(d)} (x^k - q^k) \,,
	\label{deltaN}
	\\ && 
	G^{ijkl} \nabla_{j} K_{kl}  
	=
	- \frac{ \kappa m }{ \sqrt{gL} } 
	\left( (\dot{q}^0)^2 N^i + \dot{q}^0 \dot{q}^i \right) 
	  \delta^{(d)} (x^m - q^m) \,,
	\label{deltaNi}
	\\ 
	&& 
	\frac{1}{\sqrt{g}} \frac{\partial}{\partial t} 
	  \left( \sqrt{g} G^{ijkl} K_{kl} \right) 
	+ 2 G^{mikl} \nabla_m ( N^j K_{kl} ) 
	- G^{ijkl} \nabla_{n}( N^{n} K_{kl} )  \nonumber 
	\\ &&
	+ 2 N (K^i{}_k K^{jk} - 2 \lambda KK^{ij}) 
	- \frac{1}{2} N g^{ij} G^{klmn} K_{kl} K_{mn} 
	- \beta \left( \nabla^{ij} N - g^{ij} \nabla^{2}N \right)
	\nonumber 
	\\ &&
	+ \frac{\alpha}{N} \left(\nabla^i N \nabla^j N 
	     - \frac{1}{2} g^{ij} \nabla_{k} N \nabla^{k} N \right)
	-\beta N \left( R^{ij} - \frac{1}{2} g^{ij} R \right) 
	\nonumber 
	\\ &&
	=
	\frac{ \kappa m}{ \sqrt{gL}}  
	\left( \dot{q}^{i} \dot{q}^{j} -(\dot{q}^{0})^{2} N^{i} N^{j}  \right) \delta^{(d)} (x^m - q^m) \,,
	\label{deltag}
\end{eqnarray}
where $G^{jikl} \equiv \frac{1}{2} \left( g^{ik} g^{jl} + g^{il} g^{jk} \right) - \lambda g^{ij} g^{kl}$. The equations of motion corresponding to the variations of the coordinates of the particle are
\begin{eqnarray}
	{q}^i{}'' + \Gamma^{i}_{kl} {q}^k{}' {q}^l{}' 
	+ 2 g^{il} \partial_{[k}N_{l]} {q}^{0}{}' {q}^{k}{}'  
	+ N^i {q}^{0}{}'' &&
	\nonumber
	\\
	+ \frac{1}{2} \partial^i \left( N^2 - N_k N^k \right) 
	   ( {q}^0{}' )^2
	&=& 0 \,,
	\label{partspace}
	\\
	\frac{2}{\sqrt{L}} \frac{d}{d t} 
	\left( ( N^{2} - N_{k} N^{k} ) {q}^{0}{}' - N_{k} {q}^{k}{}'\right) 
	+ \partial_0 \left( N^2 - N_k N^k \right) ({q}^0{}')^2
	&&
	\nonumber 
	\\ 
	- \partial_{0} g_{ij} {q}^{i}{}' {q}^{j}{}'
	- 2 \partial_{0} N_{k} {q}^{0}{}' {q}^{k}{}' 	
	 &=& 0 \,,
	\label{parttime}
\end{eqnarray}
where the prime means
\begin{equation}
 {\psi}' \equiv \frac{1}{\sqrt{L}}\frac{\partial \psi}{\partial t} \,. 
\end{equation}
	
\section{Gravitational field of the particle at rest}
We now consider the particle at rest located at the origin of the coordinate system of the two-dimensional spatial slices. This is achieved by fixing $q^0 = t$, $q^i = 0$. On the side of the Ho\v{r}ava gravity we impose the gauge $N_i = 0$. As we have commented, since the particle is at rest, a natural choice is to regard the fields  $g_{ij}$ and $N$ as static. As in $2+1$ General Relativity, we assume that $N$ has a fixed value at infinity, say $N_\infty = 1$. Under these settings the field equation (\ref{deltaNi}) is automatically solved. Since we are in two spatial dimensions we have the identity $R_{ij} - \frac{1}{2} g_{ij} R = 0$. Equation (\ref{deltag}) then takes the form
\begin{equation}
   	\beta \left( \nabla_{ij} N - g_{ij} \nabla^{2}N \right)
  	- \frac{\alpha}{N} \left(\nabla_i N \nabla_j N 
 	- \frac{1}{2} g_{ij} \nabla_{k} N \nabla^{k} N \right)
  	= 0 \,.
\end{equation}
The trace of this equation, assuming $\beta \neq 0$, yields
\begin{equation}
	\nabla^2 N=0 \,.
	\label{nablaN}
\end{equation}
Thus, with the given boundary condition, the lapse function is everywhere constant, $N=1$. With the information we have so far, the equations of motion of the particle (\ref{partspace}) and (\ref{parttime}) are completely solved. There remains the constraint (\ref{deltaN}) as the unique equation to be solved. It becomes	
\begin{equation}
 \sqrt{g} R = \frac{2 \kappa m }{\beta} \delta^{(2)}(x^i) \,.
 \label{riccidelta}	 
\end{equation}	
This is exactly the same equation that arises in $2+1$ General Relativity coupled to a point particle at rest, except for the coupling constant. In $2+1$ General Relativity the analogous equations is  $\sqrt{g} R = 16 \pi G_{\tiny N} m \delta^{(2)}(x^i)$, hence the coupling constant for the three-dimensional Ho\v{r}ava theory analogous to the Newton constant is $\kappa / 8 \pi \beta$. Therefore, the solution of Eq.~(\ref{riccidelta}) is functionally the same of $2+1$ General Relativity. Since this one of the central results of this paper, we reproduce here how the solution of Eq.~(\ref{riccidelta}) is found. Actually, it was found as a problem of many particles in Ref.~\cite{Deser:1983tn}, whereas the two-body form was first obtained in Ref.~\cite{Staruszkiewivcz} by a different argument. Following \cite{Deser:1983tn}, we write the spatial metric in a coordinate system where it gets an explicit conformally flat form,
\begin{equation}
 ds_{(2)}^{2} = \Omega(x,y)^{2}( dx^2 +  dy^2 ) \,.
\end{equation}  
The Ricci scalar becomes $\sqrt{g}R = - 2 \Delta \ln\Omega$, 
where $\Delta \equiv \partial_{xx} + \partial_{yy}$ is the flat Laplacian. Then Eq.~(\ref{riccidelta}) takes the form
\begin{equation}
	\Delta \ln\Omega = 
	- \frac{ \kappa m }{\beta} \delta^{(2)}(x^m) \,.
	\label{green}
\end{equation}
The two-dimensional Green function, $\Delta G(\vec{x},\vec{x}\,') = 2\pi \delta(\vec{x} - \vec{x}\,')$, has the form 
\begin{equation}
 G(\vec{x},\vec{x}\,') = 
 \ln \left( \frac{ | \vec{x} - \vec{x}\,' | }{r_0} \right) \,,
\end{equation}
where $r_0$ is an arbitrary constant. Thus, the solution of Eq.~(\ref{green}) is $ \Omega = r^{ -\frac{ \kappa m}{\pi\beta} }$,
where $r \equiv \sqrt{ x^2 + y^2 }$, and we have absorbed $r_0$ by means of a coordinate rescaling. Thus, we have that the final  solution can be expressed in polar coordinates,
\begin{equation}
	ds^{2} = 
	r^{-\frac{ \kappa m}{\pi\beta}} ( dr^2 + r^2 d\theta^2 ) \,. 
	\label{polar}
\end{equation}
For a better understanding of the geometry the following coordinate transformation is useful,
\begin{equation}
	\rho = \frac{1}{\gamma} r^{\gamma} \,,
	\quad
	\theta\,' = \gamma \theta \,,
	\quad
	\gamma \equiv 1 - \frac{ \kappa m}{2 \pi \beta} \,.
\end{equation}
The line element (\ref{polar}) acquires an explicit flat form,
\begin{equation}
	ds^2 = d\rho^2 + \rho^2 d{\theta\,'}^2 \,.
\end{equation}
If $m = 0$ the solution is a regular, globally flat geometry. For nonzero values of $m$ there are three qualitatively different kinds of geometries, depending on the three cases: $m < 2\pi \beta / \kappa $, $m > 2\pi \beta / \kappa$ and $m = 2\pi \beta / \kappa$, which give $\gamma > 0$, $\gamma < 0$ and $\gamma = 0$ respectively. For $\gamma > 0$ the geometry is a flat cone with a singularity at the origin, where the massive particle is located. The points with infinite $r$ are at infinite proper distance from any point of the interior of the cone \cite{Ashtekar:1993ds}. This is a requisite for the expected asymptotic-flatness condition defined on the basis of this solution. The range of the cone coordinates are $\rho \in [0,\infty)$ and $\theta\,' \in [0,2\pi\gamma]$. Outside the singularity the geometry is flat. The cone can have a deficit angle or an excess angle depending on whether $\kappa m / \beta$ is positive or negative respectively. The presence of the coupling constant $\beta$ allows for more possibilities of sign than in General Relativity. Even in the $3+1$ Ho\v{r}ava theory one has an argument for restricting the sign of $\beta$: $\sqrt{\beta}$ is the speed of the tensorial modes. Below we will see that the speed of the unique propagating mode in the $2+1$ theory does depend on $\beta$, but it does not restrict its sign. Therefore, with the information we have found in this paper we do not get restrictions on the sign of $\beta$. For $\gamma < 0$ the original location of the particle, $r = 0$, lies now at infinity in the coordinate $\rho$, whereas the infinite distance from the particle in the $r$ coordinate is now at finite proper distance of the interior points. In this case the space curls up. For $\gamma = 0$ the geometry corresponds to a cylinder. The two last geometries are not compatible with the notion of asymptotic flatness, they are actually excluded. Further discussion on these two cases can be found in Refs.~\cite{Deser:1983tn,Ashtekar:1993ds}.

\section{Asymptotic flatness}
In $2+1$ General Relativity the asymptotic flatness of the gravitational field variables is managed by taking the point-particle solution as  giving the dominant mode in the asymptotic expansion \cite{Ashtekar:1993ds}. Here we adopt the same point of view, guided by the exact solution of the previous section. Since the solution is the same of General Relativity, the asymptotically flat condition for the Ho\v{r}ava theory we postulate is also the same, hence our exposition is parallel to the known case of General Relativity. 

In the asymptotic region of the spatial slices one can introduce polar coordinates $r,\theta$ and their corresponding Cartesian coordinates $x,y$. We assume that if a variable is of order $\mathcal{O}(r^{n})$ asymptotically, then its derivatives $\partial_i \mathcal{O}(r^{n})\sim\mathcal{O}(r^{n-1})$. The lapse function approaches its fixed value according to
\begin{equation}
N = 1 + \mathcal{O}(r^{-1}) \,.
\label{NN}
\end{equation}
The asymptotic form of the metric in two spatial dimensions in Cartesian coordinates is given by
\begin{equation}
	g_{ij} = 
	r^{-\mu} ( \delta_{ij} + \mathcal{O}(r^{-1}) ) \,,
	\label{metricasin}
\end{equation}
where $\mu$ is an arbitrary constant. It is useful to write an arbitrary variation of the metric (\ref{metricasin}), considering that the value of the constant $\mu$ varies among the different configurations,
\begin{equation}
	\delta g_{ij}
	= r^{-\mu} (- \delta \mu \ln r\, \delta_{ij} 
	+ \mathcal{O}(r^{-1})) \,.
\end{equation}
Other composed objects that we need are
\begin{equation}
	R \sim \mathcal{O}( r^{\mu - 3} ) \,,
	\quad
	a_i \sim \mathcal{O}(r^{-2}) \,.
\end{equation}
For the canonical momentum $\pi^{ij}$ we follow a criterium analogous to the one used in Ref.~\cite{Ashtekar:1993ds} for $2+1$ General Relativity. On basic grounds, the kinetic term of the canonical action must be well defined, hence the integral
\begin{equation}
 \int d^2x \pi^{ij} \dot{g}_{ij}
\end{equation}
must converge. This requisite is satisfied if we assume that the asymptotic behavior of the canonical momentum is
\begin{equation}
\pi^{ij} \sim 
\mathcal{O} (r^{\mu - 2}) \,.
\end{equation}

\section{Canonical formulation}
Our aim in this section is to perform first the general Hamiltonian formulation and then study the conditions for the differentiability of the Hamiltonian and the relationship with the energy. For the sake of completeness, in the first part we keep the coupling to the relativistic particle.

\subsection{General formalism}
We follow the route of obtaining the Hamiltonian from a Legendre transformation of the action (\ref{action}). Our approach is close to the approach of Ref.~\cite{Menotti:1999pn}, which is devoted to the system General Relativity-point particle in $2+1$ dimensions. In order to simplify the notation, in this section we use units such that $2 \kappa = 1$, and to simplify the computations we set $q^0(t) = t$.  The canonical gravitational and particle momenta are obtained by the Legendre transformation,
\begin{equation}
\pi^{ij} = \sqrt{g} ( K^{ij} - \lambda g^{ij}K ) \,, 
\quad
p_i = \frac{m}{\sqrt{L}} \left( g_{ij} \dot{q}^j + N_i (q)   \right) \,.
\end{equation}
The total Hamiltonian of the system Ho\v{r}ava gravity-point particle in $d$ spatial dimensions is
\begin{eqnarray}
 H &=&
 \int d^dx 
 \left[ \frac{N}{\sqrt{g}} \left( \pi^{kl} \pi_{kl} 
   + \frac{ \lambda}{ 1 - \lambda d} \pi^2 \right)
   - \sqrt{g} N \left( \beta R + \alpha a_k a^k \right)
   + N_k \mathcal{H}^{k} + \sigma P_{N} \right]   
 \nonumber
 \\ &&
   + N(q) \sqrt{ p_k p^k + m^2 } - N_{k}(q) p^{k}  \,,
 \label{hamiltonian}
\end{eqnarray} 
$\mathcal{H}^i$ is the sourced momentum constraint,
\begin{equation}
\mathcal{H}^{i} =
- 2 \nabla_j \pi^{ij} - p^i \delta^{(d)}(x^k - q^k) 
= 0 \,,
\label{Hk}
\end{equation}
and the shift vector $N_i$ plays the role of its Lagrange multiplier. $P_N$ is the momentum canonically conjugated to $N$, which is zero. We have incorporated this primary constraint to the Hamiltonian with the Lagrange multiplier $\sigma$. In all the canonical formalism we assume $\lambda \neq 1/d$.

In $d=2$ dimensions, the terms of the Hamiltonian that depend on the canonical momentum, 
\begin{equation}
\int dx^2 \frac{N}{\sqrt{g}} 
\left( \pi^{kl} \pi_{kl} + \frac{\lambda}{ 1 - 2 \lambda } \pi^2  \right)  \,,
\label{kc}
\end{equation}
impose a bound on the asymptotic flatness condition, like in $2+1$ General Relativity, which in turn becomes a bound on the value of the Hamiltonian \cite{Ashtekar:1993ds}. This comes from the requisite that the operator (\ref{kc}) is well defined, since it should be the generator of the time evolution of the spatial metric, that is, the variation of the Hamiltonian should yield the equation for $\dot{g}_{ij}$. This implies that the integral in (\ref{kc}) must converge. Since the integrand has an asymptotic power of $r^{\mu - 4}$, the integral does not converge unless $\mu < 2$.

The time preservation of the $P_N = 0$ constraint leads to the sourced Hamiltonian constraint
\begin{equation}
\begin{array}{rcl}
	\mathcal{H} &\equiv& 
	{\displaystyle
	\frac{1}{\sqrt{g}} \left( \pi^{kl} \pi_{kl} 
	+ \frac{\lambda}{1-\lambda d} \pi^2 \right) 
	- \sqrt{g} \left[ \beta R 
	- \alpha \left( a_k a^k + 2 \nabla_{k}a^{k} \right) \right]
	}
	\\[2ex] &&
	{\displaystyle
	+ \sqrt{ p_k p^k + m^2 } \delta^{(d)}(x^i - q^i) 
	= 0 \,. }
\end{array}
	\label{H}
\end{equation}
In turn, the preservation of the Hamiltonian constraint leads to a partial differential equation for the Lagrangian multiplier $\sigma$. Dirac's procedure ends with this step. The theory has the constraints $P_N = 0$, $\mathcal{H}^i=0$ and $\mathcal{H}=0$. The balance between field variables, constraints and gauge symmetries indicates that the theory propagates a single (even) degree of freedom.
	
The equations of motion in the canonical formalism in $d$ spatial dimensions are
\begin{eqnarray}
	\dot{g}_{ij} &=&
	\frac{2N}{\sqrt{g}} \left( \pi_{ij} 
	+ \frac{\lambda}{1 - \lambda d} g_{ij} \pi \right)
	+ 2\nabla_{(i} N_{j)} \,,
	\label{dotg}
	\\
	\dot{\pi}^{ij} &=&
	- \frac{2N}{\sqrt{g}} \left[ \pi^{ik} \pi^{j}{}_{k}
	  - \frac{1}{4} g^{ij} \pi^{kl} \pi_{kl}
	   +\frac{\lambda}{ 1 - \lambda d}  
	   \left( \pi \pi^{ij} - \frac{1}{4} g^{ij} \pi^2  \right)
	    \right]	
	\nonumber 
	\\ &&
	- \beta N \sqrt{g} \left( R^{ij} - \frac{1}{2} g^{ij} R \right)
	- \alpha \frac{\sqrt{g}}{N} \left( \nabla^i N \nabla^j N 
	              - \frac{1}{2} g^{ij} \nabla_k N \nabla^k N \right)
	\nonumber 
	\\ &&
	+ \beta \sqrt{g} \left( \nabla^{ij} N - g^{ij} \nabla^2 N \right) 
	- 2 \sqrt{g} \nabla_k ( N N^{(i} \pi^{j)k} )
	\nonumber 
	\\ &&
	+ \sqrt{g} \nabla^k ( N N_k \pi^{ij} )
	- \left( N^i N^j 
	 - \frac{ p^i p^j }{ 2 \sqrt{ p_k p^k + m^2}} \right)   
	     \delta^{(d)}(x^i - q^i) \,,
	\label{dotpi}
	\\
	\dot{q}^i &=&
	- N^i(q) + \frac{N(q) p^i}{ \sqrt{ p_k p^k + m^2} } \,,
	\\
	\dot{p}_i &=& 
	\partial_i \left( N_k(q) p^k - N(q) \sqrt{ p_k p^k + m^2} \right) \,.
\end{eqnarray}

\subsection{Linearized equations}
We briefly confirm, by means of a linear-order perturbative analysis, that the (sourceless) theory propagates one scalar degree of freedom.  This provides completeness to our analysis and allow to test the consistency of the Hamiltonian formulation. The perturbative analysis around a globally flat space, which is the background with $\mu = 0$, such that the conical singularity disappears, was already done for the theory in the Lagrangian formalism with all relevant operators in Ref.~\cite{Sotiriou:2011dr}. We also consider the perturbations around the globally flat space, hence we set $\mu = 0$ for this analysis. However, other background geometries are possible for other values of $\mu$, that is, one in principle can study perturbations around a cone.

The canonical variables $g_{ij},\pi^{ij}$ and $N$ are perturbated according to
\begin{equation}
 g_{ij} = \delta_{ij} + h_{ij} \,,
 \quad
 \pi^{ij} = p_{ij} \,,
 \quad
 N = 1 + n \,,
\end{equation}
and the Lagrange multiplier $N_i = n_i$. Since we are working in the canonical formalism, we introduce the transverse/longitudinal decomposition in the two spatial dimensions, which allows us to solve the momentum constraint (\ref{Hk}). With the given boundary conditions on the canonical variables the flat Laplacian $\Delta$ is invertible. We introduce the two-dimensional decomposition
\begin{equation}
 h_{ij} = 
 \left( \delta_{ij} - \frac{\partial_{ij}}{\Delta} \right) h^T
 + \partial_{(i} h^L_{j)} \,,
\end{equation}
and similarly for $p_{ij}$. We fix the gauge symmetry of two-dimensional spatial diffeomorphisms by imposing the transverse gauge, $h^L_i = 0$. The linear-order version of the momentum constraint (\ref{Hk}) takes the form
\begin{equation}
 \partial_i p_{ij}= 0 \,.
\end{equation}
This is equivalent to $p^L_i=0$. The linear-order version of the constraint $\mathcal{H}$ (\ref{H}) takes the form
\begin{equation}
 \beta \Delta h^T + 2 \alpha \Delta n = 0 \,.
\end{equation}
Assuming $\alpha \neq 0$, this equation implies
\begin{equation}
 n = - \frac{\beta}{2\alpha} h^T \,.
 \label{soln}
\end{equation}
Therefore, the constraints $\mathcal{H}^i$ and $\mathcal{H}$ and the transverse gauge fix the variables $h^L_i$, $p^L_i$ and $n$, leaving the transverse sector $h^T,p^T$ and the Lagrange multiplier $n_i$ active. Now we move to the canonical equations of motion (\ref{dotg}) and (\ref{dotpi}). The longitudinal sector of Eq.~(\ref{dotg}) yields an equation for $n_i$,
\begin{equation}
 \Delta n_i + \partial_{ij} n_j = 
 -\frac{2\lambda}{ 1 - 2\lambda } \partial_i p^T \,,
\end{equation}
whose solution is
\begin{equation}
 n_i = 
 - \frac{ \lambda }{ 1 - 2 \lambda } 
    \partial_i \left( \frac{1}{\Delta} p^T \right) \,.
 \label{ni}
\end{equation}
Note that in Ho\v{r}ava gravity, unlike General Relativity, we do not have the freedom to set, for example, $t = \frac{1}{\Delta} p^T$. In the generic formulation of the Ho\v{r}ava gravity there is no gauge symmetry, additional to the spatial diffeomorphisms, that allows to obtain $n_i = 0$. This variable can be indeed switched off by a gauge choice, as we did in section 3, but using precisely the spatial diffeomorphisms symmetry, and in this perturbative analysis we have already use them to impose the transverse gauge. A different situation happens in the kinetic-conformal formulation, where the condition $p^T = 0$ enter in the game as a constraint of the theory. In this case, by an analysis parallel to the one leading to Eq.~(\ref{ni}), one obtains $n_i = 0$. The trace of the linearized Eqs.~(\ref{dotg}) and (\ref{dotpi}) leads automatically to their transverse sectors. The trace of these equations, after using (\ref{soln}) yield, respectively,
\begin{equation}
\dot{h}^T = 2 \left( \frac{1 - \lambda }{1-2\lambda} \right) p^T \,,
\quad
\dot{p}^T  = \frac{\beta^2}{2\alpha} \Delta h^T \,.
\end{equation}
These equations imply
\begin{equation}
 \ddot{h}^T - \frac{\beta^2}{\alpha} \left(\frac{1 - \lambda}{1-2\lambda} \right) 
 	\Delta h^T = 0 \,.
\end{equation}
Thus, the transverse scalar $(h^T, p^T)$ is the physical mode propagated by the theory, with squared speed $\beta^2 (1 - \lambda) / (1-2\lambda)\alpha$ at the level of the linearized theory. As we commented previously, the speed of the scalar gravitational wave in the $z=1$ theory depends on the absolute value of $\beta$. The stability of this mode, in the effective theory, impose the bound
\begin{equation}
 \frac{1}{ \alpha} \left(\frac{ 1 - \lambda }{1 - 2\lambda}\right) > 0 \,.
\end{equation}
This bound was found in Ref.~\cite{Sotiriou:2011dr}.

\subsection{Differentiability of the Hamiltonian and the energy}
Similarly to the case of General Relativity, one may ask whether the asymptotic conditions impose restrictions on the differentiability of the Hamiltonian. For this aspect we study the purely gravitational theory, without coupling to sources. We remark that in the $z=1$ potential of the Hamiltonian, 
\begin{equation}
 - \sqrt{g} N \left( \beta R + \alpha a_k a^k \right) \,,
\end{equation}
besides the Ricci-scalar term, we have the $a^2$ term characteristic of the nonprojectable Ho\v{r}ava theory. The variation of the Ricci-scalar term with respect to the metric yields	
\begin{equation}
	-\beta \delta \int dx^2 \sqrt{g} N R =
	-\beta \int dx^2 \sqrt{g} 
	\left( - \nabla_{ij} N + g_{ij} \nabla^2 N \right) \delta g^{ij} 
	- \beta \delta \mu \oint d\theta \,,
	\label{varg}
\end{equation}
where we have used the fact that the Einstein tensor vanishes in two dimensions. The second term in the right-hand side of (\ref{varg}) is a boundary term posing an obstruction to the differentiability of the Hamiltonian, hence we must add the counterterm
\begin{equation}
  +2\pi \beta \mu \,.
\end{equation}
Of course, this is the same counterterm of $2+1$ General Relativity \cite{Ashtekar:1993ds}, except for the presence of the coupling constant $\beta$ (recalling that we use units with $2\kappa = 1$, otherwise $\kappa$ also arises in the counterterm). The dependence on $N$ of the Ricci-scalar term is algebraic, hence it gives no total derivatives when $N$ is variated. 

The $a^2$ term is algebraic in the spatial metric, hence its variation with respect to the metric yields no total derivative. Its variation with respect to $N$ yields
\begin{equation}
 - \alpha \delta \int d^2x \sqrt{g} N a_k a^k =
 \alpha \int d^2x \sqrt{g} 
    \left( a_k a^k + 2 \nabla_k a ^k \right) \delta N
 - 2\alpha \int d^2x \sqrt{g} \nabla_k \left( a^k \delta N \right) \,.
\end{equation}
The last term is equal to a line integral at the infinite boundary,
\begin{equation}
 - 2 \alpha \oint\limits_\infty dl n_k a^k \delta N \,.
 \label{oint}
\end{equation}
According to the asymptotic flatness condition, we have that the orders of these objects are $dl \sim r^{-\frac{\mu}{2} + 1}$, $n_k \sim r^{-\frac{\mu}{2}}$, $a^k \sim r^{\mu - 2}$ and $\delta N \sim r^{-1}$. Therefore the integrand decays as $r^{-2}$, such that the whole integral is zero. Thus, in $2+1$ dimensions, the $a^2$ term is differentiable with respect to $N$ under the asymptotically flat conditions we have considered, with no need of boundary counterterms.

Although the Hamiltonian (\ref{hamiltonian}) is apparently given by a nonzero integral in the two-dimensional spatial slices, it can be actually written as a sum of constraints. This is similar to the nonprojectable Ho\v{r}ava theory in $3+1$ dimensions \cite{Donnelly:2011df}, but with the difference that there is no remaining boundary term in the procedure (the boundary term that indeed remains is the one seen in Eq.~(\ref{varg}), but this is a different thing). The difference between the bulk part of the Hamiltonian, 
\begin{equation}
H_{\mbox{\tiny bulk}} =
\int d^2x 
\left[ \frac{N}{\sqrt{g}} \left( \pi^{kl} \pi_{kl} 
+ \frac{ \lambda}{ 1 - \lambda d} \pi^2 \right)
- \sqrt{g} N \left( \beta R + \alpha a_k a^k \right) \right]   
\,,
\label{bulkhamiltonian}
\end{equation} 
and the integral of the constraint $\mathcal{H}$ given in (\ref{H}),
\begin{equation}
 \int d^2x N \mathcal{H} \,,
 \label{intnh}
\end{equation}
is the integral of a total divergence,
\begin{equation}
 2 \alpha \int d^2x \sqrt{g} \nabla_k ( N a^k ) 
 = 2 \alpha \oint\limits_{\infty} dl n_k N a^k
 \,. 
 \label{divergence}
\end{equation}
In this boundary integral the balance is similar to the one we did in Eq.~(\ref{oint}), in this case with $N$ constant at infinity. The integrand decays as $r^{-1}$, hence the integral (\ref{divergence}) is zero. Therefore, the bulk part of the Hamiltonian (\ref{bulkhamiltonian}) is identical to the integral of the constraint (\ref{intnh}). We may write the full Hamiltonian as a sum of constraints plus the boundary counterterm needed for its differentiability,
\begin{equation}
	H =
	\int d^Dx 
	\left( N \mathcal{H} + N_k \mathcal{H}^k + \sigma P_N \right) + E \,,
	\label{HND}
\end{equation}
where the boundary term is given by $E = 2\pi \beta \mu$. This result implies that the energy of the gravitational field of the $2+1$ nonprojectable Ho\v{r}ava theory is given by $E$, with $\mu$ varying among the different solutions. The constant $\mu$ affects the dominant mode in the asymptotic expansion since it determines the power of asymptotic growning/decay in the coordinate system used in (\ref{metricasin}). This is the same behavior of $2+1$ General Relativity. Again, the difference is the presence of the coupling constant $\beta$ ($E = \pi \beta \mu / \kappa$ in other units).


\section{Conclusions}
Our main conclusion is the rather surprising result that the solution for the gravitational field of a massive point particle at rest in $2+1$ nonprojectable Ho\v{r}ava gravity is exactly the same solution of $2+1$ General Relativity, except for the different coupling constant that arises in Ho\v{r}ava theory. The solution is a flat cone with a deficit or excess angle proportional to the mass of the particle \cite{Deser:1983tn}. It was found in the large-distance effective action for the Ho\v{r}ava theory, which is a nonrelativistic theory. To arrive at this solution in the Ho\v{r}ava theory we have assumed static lapse function and spatial metric, since this is a natural setting for the gravitational field of the particle at rest, but the proof that the staticity condition is the only admissible condition for the system Ho\v{r}ava gravity - point particle at rest has not been given yet. 

On the basis of this exact solution, which represents the gravitational field of the simplest isolated body, we have defined the condition for asymptotic flatness exactly in the same way of $2+1$ General Relativity \cite{Ashtekar:1993ds}. An outstanding feature of the three-dimensional scenario is the functional variability of the  dominant mode in the asymptotic expansion.

We have also checked that the Hamiltonian is differentiable after the same boundary counterterm of General Relativity is added to it. Since the Hamiltonian eventually gets the form of a sum of constraints plus the boundary term, this boundary term gives the energy in the asymptotically flat case. This energy is the same of $2+1$ asymptotically flat General Relativity. This coincidence is also relevant since one could naively expect that the $(\partial_i \ln N)^2$ term of the $z=1$ action contribute to the energy, as in the $3+1$ case, but it does not. Another consequence is that for the $2+1$ nonprojectable Ho\v{r}ava theory there arise the same upper bound on the value of the Hamiltonian that was found in Ref.~\cite{Ashtekar:1993ds} for $2+1$ General Relativity.

The $2+1$ nonprojectable Ho\v{r}ava theory has the fundamental difference with $2+1$ General Relativity that it is not a topological theory, instead it propagates an even scalar mode. However, some features are qualitatively the same between both theories, as we have seen here for the case of the point-particle solution and the associated asymptotic flatness. If we put together these two facts, some interesting possibilities arise. For example, gravitational waves propagating on a cone can be supported by the three-dimensional Ho\v{r}ava theory. We hope several studies like this one can be further developed.



\end{document}